\documentclass[12pt]{iopart}
\usepackage{iopams}  

\usepackage{graphicx}
\usepackage{bm}

\newcommand{\be}{\begin{equation}}
\newcommand{\ee}{\end{equation}}
\newcommand{\eq}[1]{Eq.~(\ref{#1})}
\newcommand{\fig}[1]{Fig.~\ref{#1}}
\def\bea{\begin{eqnarray}}
\def\eea{\end{eqnarray}}

\def\bra{\langle}
\def\ket{\rangle}
\def\vq{{\bf q}}

\def\vk{{\bf k}}

\begin{document}

\title[Ferromagnetic and metamagnetic transitions in itinerant electron systems]{Ferromagnetic and metamagnetic transitions in itinerant electron systems: a microscopic study}

\author{Hiroyuki Yamase}

\address{International Center of Materials Nanoarchitectonics, 
National Institute for Materials Science, Tsukuba 305-0047, Japan }
\ead{yamase.hiroyuki@nims.go.jp}
\vspace{10pt}
\begin{indented}
\item[]December 8, 2022
\end{indented}

\begin{abstract}
We perform a microscopic study of itinerant ferromagnetic systems. We  reveal a very rich phase diagram in the three-dimensional space spanned by the chemical potential, a magnetic field, and temperature beyond the Landau theory analyzed so far.  Besides a  generic wing structure near a tricritical point upon introducing the magnetic field, we find that an additional wing can be generated close to a quantum critical end point (QCEP) and also even from deeply inside the ferromagnetic phase. A tilting of the wing controls the entropy jump associated with the metamagnetic transition. Ferromagnetic and metamagnetic transitions are usually accompanied by a Lifshitz transition at low temperatures, i.e., a change of Fermi surface topology including the disappearance of the Fermi surface.  In particular, the Fermi surface of either spin band vanishes at the QCEP. These rich phase diagrams are understood in terms of the density of states and the breaking of particle-hole symmetry in the presence of a next nearest-neighbor-hopping integral $t'$, which is expected in actual materials. The obtained phase diagrams are discussed in a possible connection to itinerant ferromagnetic systems such as UGe$_2$, UCoAl, ZrZn$_2$, and others including materials exhibiting the magnetocaloric effect.  
\end{abstract}

%
%
%
%
%

\section{Introduction}
Itinerant metamagnetism is known as a magnetic phase transition accompanied by a discontinuous change of 
magnetization upon applying a magnetic field \cite{levitin89}. 
Typical examples are intermetallic compounds such as ${\rm Co(Se_{1-x}S_{x})_{2}}$ \cite{adachi79}, 
${\rm Y(Co_{1-x}Al_{x})_{2}}$ \cite{aleksandryan85,sakakibara86,goto89}, 
${\rm Lu(Co_{1-x}Al_{x})_{2}}$ \cite{gabelko87,fukamichi01}, 
and giant magnetocaloric effect materials such as MnAs \cite{gama04}, ${\rm Gd_{5}(Si_{1-x}Ge_{x})_{4}}$ \cite{tegus02a,tegus02b}, 
and ${\rm MnFeP_{1-x}As_{x}}$ \cite{tegus02a,tegus02b}. 
Theoretical studies of itinerant metamagnetism have a long history \cite{wohlfarth62} 
including an issue of first-order ferromagnetic instability \cite{lidiard51,bean62,shimizu64}. 
As summarized in Refs.~\cite{levitin89} and \cite{shimizu82}, the itinerant metamagnetic transition is expected 
when the density of states (DOS) $N(\epsilon)$ exhibits a positive curvature 
$(\frac{d^2 N}{d \epsilon^2} > 0)$ near the Fermi energy. 
This condition is easily understood in terms of the Landau free energy \cite{shimizu82}. 
The free energy of the system may be expanded as 
\be
F(m) = a_2 m^2 + a_4 m^4 + a_6 m^6 - h m \,,
\label{Landau}
\ee
where $m$ is magnetization and $h$ is a magnetic field. 
The large DOS usually leads to a relatively small value of $a_2 (>0)$ and the positive curvature of the DOS 
can yield a negative value of $a_4$. In general, it is possible to find a parameter region, where $a_6$ is positive. 
Hence the Landau free energy with $h=0$ can have two local minima at $m=m_1(=0)$ and $m_2 (> m_1)$ in a positive region of $m$.  
We suppose a situation of $F(m_1) < F(m_2)$, i.e., a paramagnetic phase. 
In this case, upon applying a field $h$, $F(m_1) $ becomes lower with a finite $m_1$, but $F(m_{2})$ is 
lowered more substantially. Eventually 
the level crossing of the two local minima occurs, leading to a first-order transition 
from a state with $m_1$ to that with $m_2$, which describes a metamagnetic transition. 

This basic concept was developed by considering the renormalization of the coefficients $a_2$, $a_4$, and $a_6$ in \eq{Landau} 
by magnetic fluctuations \cite{yamada93}, where the temperature dependence of the bare coefficients was discarded. 
In particular, a phase diagram was constructed by using the renormalized coefficients of 
the Landau free energy \eq{Landau} \cite{goto01,yamada03}. 
The ferromagnetic phase transition is second order at high temperatures and 
changes to a first order at low temperatures, whereas the metamagnetic transition takes place above 
the first-order transition temperature by applying a magnetic field.  
Detailed comparisons with experiments were made for various metamagnetic materials such as 
${\rm Co(Se_{1-x}S_{x})_{2}}$ \cite{goto01,yamada03,goto97},  
${\rm Y(Co_{1-x}Al_{x})_{2}}$ \cite{goto01,yamada03,bloch75,takahashi95},  
${\rm Lu(Co_{1-x}Ga_{x})_{2}}$ \cite{goto01,yamada03}, 
${\rm La(Fe_{x}Si_{1-x})_{13}}$ \cite{yamada03},  
UCoAl \cite{goto01}, 
${\rm MnFeP_{1-x}As_{x}}$ \cite{yamada03}, and 
UGe$_{2}$ \cite{kabeya10}, and a qualitatively good agreement was concluded.

A different scenario was also proposed. 
Reference~\cite{belitz99} showed that itinerant ferromagnets generally exhibit 
a first-order phase transition at low temperatures without invoking a large DOS. 
In a metallic system, there are always gapless particle-hole excitations around the Fermi surface. 
These excitations yield a nonanalytic term in the free energy. 
Such a term in turn generates a local minimum in the potential and can lead to a first-order transition 
at low temperatures. The transition is expected to change to a continuous one at high temperatures because  
the singular term is cut off by temperature. 
The resulting phase diagram shares many features with the conventional theory \cite{goto01,yamada03}. 
A similar nonanalytic term was also obtained in Fermi liquid theory 
and was invoked to explain a metamagnetic transition \cite{misawa94}. On the other hand, 
it was shown that the nonanalytic term preempts the first-order ferromagnetic transition 
and leads to inhomogeneous states \cite{karahasanovic12,conduit13}.

The end point of a second-order transition line corresponds to a tricritical point (TCP), 
at which the transition becomes first order. 
As is well known in theory of the TCP \cite{griffiths70,lawrie84}, first-order transition {\it surfaces} 
develop by applying a magnetic field 
and three critical lines merge together at the TCP: 
one is the second-order transition line at zero field and 
two are critical end lines (CELs) corresponding to the edges of the first-order transition surfaces  in a finite field. 
A metamagnetic transition occurs when the system crosses the first-order transition surfaces by tuning the field. 

Experimentally, a phase diagram similar to the general phase diagram associated with the TCP 
was observed in  itinerant ferromagnetic materials such as UGe$_2$ \cite{taufour10,kotegawa11},   
UCoAl \cite{aoki11}, and ZrZn$_{2}$ \cite{kimura04,uhlarz04}.  
On one hand, this observation supports the phase diagram \cite{belitz05} obtained in the mechanism 
in terms of the nonanalytic term  \cite{belitz99}. 
However, we shall show in the present paper that a phase digram similar to 
experimental observations can also be captured 
by considering the large DOS with a positive curvature in line with early theoretical studies \cite{levitin89,wohlfarth62,lidiard51,bean62,shimizu64,shimizu82,yamada93,goto01,yamada03,goto97}.  

We also notice that many theoretical studies were performed phenomenologically in detail  \cite{levitin89,wohlfarth62,lidiard51,bean62,shimizu64,shimizu82,yamada93,goto01,yamada03,goto97,belitz05,yamada07,berridge10}, but microscopic studies of the itinerant metamagnetic transition are rather sparse    \cite{sandeman03,berridge11,wysokinski15}. In addition, the majority of the earlier studies \cite{levitin89,shimizu82,yamada93,goto01,yamada03,goto97,belitz05,yamada07,berridge10} is based on Landau expansion, which is valid only for a small magnetization. However, a metamagnetic transition is accompanied by a large change of the magnetization and it is not clear how reliable those works are, requiring a microscopic analysis without Landau expansion. Moreover, a giant magnetocaloric effect is currently an important topic aiming to  develop magnetic refrigeration \cite{franco18}, which will achieve high energy efficiency and will be used for hydrogen liquefaction \cite{numazawa14}. In this respect, a metamagnetic phase transition attracts much interest because of a possibly considerable entropy change \cite{terada21}. Hence it should be also valuable to establish a canonical phase diagram of itinerant ferromagnetism as well as metamagnetism. 

In this paper, we introduce a microscopic mean-field exact model of itinerant ferromagnetism on a square lattice in Sec.~2. We then perform comprehensive calculations to establish three-dimensional phase diagrams spanned by temperature, the chemical potential, and a magnetic field in Sec.~3, which is beyond the scope of the Landau theory analyzed previously \cite{levitin89,shimizu82,yamada93,goto01,yamada03,goto97}. 
The key advantage of the present microscopic analysis over phenomenological studies \cite{wohlfarth62,lidiard51,bean62,shimizu64,belitz05,yamada07,berridge10} is that we can elucidate not only the generic wing structure associated with the metamagnetic transition near a TCP, but also additional wings developing near a quantum critical end point (QCEP) and also from deeply inside the ferromagnetic phase. On top of these important findings, the present work clarifies the microscopic origin of the wing, the microscopic reason of the occurrence of the QCEP, the relation to the entropy jump, the connection to a Lifshitz transition, and the importance of a next nearest-neighbor-hopping integral $t'$. In Sec.~4, we discuss the obtained results from both theoretical and experimental viewpoints. In particular, our obtained phase diagrams capture characteristic features observed in UGe$_{2}$ \cite{taufour10,kotegawa11}, UCoAl  \cite{aoki11}, and ZrZn$_{2}$ \cite{kimura04,uhlarz04}.

\section{Model and Formalism}
To elucidate the typical phase diagram of itinerant ferromagnetism including the effect of the Zeeman field, 
we study the following mean-field exact Hamiltonian on a square lattice: 
\be
 \mathcal{H}= \sum_{\vk, \sigma} \xi_{\vk}^{0} c_{\vk \sigma}^{\dagger} c_{\vk \sigma} 
- \frac{g}{2N} \sum_{\vk \vk' \sigma \sigma'} \sigma \sigma' 
c_{\vk \sigma}^{\dagger} c_{\vk \sigma} c_{\vk' \sigma'}^{\dagger} c_{\vk' \sigma'} 
- \frac{h}{2} \sum_{\vk \sigma} \sigma  c_{\vk \sigma}^{\dagger} c_{\vk \sigma} \,,
\label{hamiltonian}
\ee
where $c_{\vk \sigma}^{\dagger}$ and $c_{\vk \sigma}$ are the creation and annihilation operators 
for electron with momentum $\vk$ and  spin orientation $\sigma (= 1\; {\rm or}\; -1)$, respectively; 
$N$ is the total number of the lattice sites. The first term is a kinetic term of electrons and its dispersion is given by 
\be
\xi_{\vk}^{0} = - 2 t ( \cos k_x + \cos k_y) -4 t'   \cos k_x  \cos k_y -\mu  \,.
\ee
Here $t$ and $t'$ are hopping integrals between the nearest-neighbor and next 
nearest-neighbor sites, respectively, and $\mu$ is the chemical potential. 
The second term in \eq{hamiltonian} is the same as the Landau interaction function in the $s$-wave 
spin-antisymmetric channel \cite{negele}.  
It describes a spin-dependent forward scattering interaction and can lead to 
ferromagnetism for $g>0$. The interaction is considered only for the spin quantization axis parallel to the $z$-axis. 
Hence the present magnetic interaction is expected to describe an itinerant ferromagnetic system with a strong spin anisotropy 
observed in several U-based materials \cite{aoki11,aoki12}. 
Our interaction may also be applicable to a system with spin rotational invariance by considering that 
our interaction is obtained after a mean-field approximation to a certain microscopic model. 
For example, if one starts with the Hubbard interaction $U\sum_{i} c^{\dagger}_{i \uparrow} c_{i \uparrow}  c^{\dagger}_{i \downarrow} c_{i \downarrow}$ where $i$ represents the lattice site, one obtains $g=U/2$ in momentum space by assuming a ferromagnetic mean-field solution, namely by focusing on the zero-momentum transfer channel---the forward scattering channel. 
In fact, it is well recognized in the Hubbard model that the ferromagnetic interaction actually becomes dominant when $| t'/t |$ is relatively large around $0.3-0.5$ for the density around van Hove filling; otherwise, antiferromagnetism and $d$-wave superconductivity would be stabilized \cite{honerkamp01,katanin03,husemann09}. Respecting this insight, one might choose a large $t'$ in our model, although a choice of our parameters is arbitrary in \eq{hamiltonian} in principle.  Note that ferromagnetism is stabilized around van Hove filling also in the present model as we shall see in Sec.~3. 
The third term is the Zeeman energy and $h$ is an effective magnetic field defined as $h=\mathfrak{g} \mu_{B} H$, where 
$\mathfrak{g}$ is a $g$ factor, $\mu_{B}$ the Bohr magneton, and $H$ a magnetic field. 

Other terms not included in our Hamiltonian may be present in real materials. Our interest lies in the case where 
ferromagnetism becomes a leading instability. Hence the Hamiltonian, \eq{hamiltonian}, should be regarded as an effective minimal model in the low-energy region, 
which may describe a system close to ferromagnetic instability such as UGe$_{2}$ \cite{taufour10,kotegawa11,aoki12}, UCoAl \cite{aoki11}, ZrZn$_{2}$ \cite{kimura04,uhlarz04}, and UCoGe \cite{ohta10}.

The Hamiltonian, \eq{hamiltonian}, considers the electron-electron interaction only for the forward scattering 
interaction of electrons. This means that fluctuations around a mean field vanish in the limit of $N \rightarrow \infty$, 
indicating that a mean-field theory can solve the model [\eq{hamiltonian}] exactly  in the thermodynamic limit. 
We then decouple the interaction by introducing the magnetization 
\be
m=\frac{g}{2N} \sum_{\vk \sigma} \sigma \bra c_{\vk \sigma}^{\dagger} c_{\vk \sigma} \ket \,.
\ee 
The resulting mean-field Hamiltonian is given by 
\be
\mathcal{H}_{\rm MF} = \sum_{\vk \sigma} \xi_{\vk \sigma} c_{\vk \sigma}^{\dagger} c_{\vk \sigma} 
+ \frac{2N}{g} m^2 \,,
\ee
where $\xi_{\vk \sigma} = \xi_{\vk}^{0} - \frac{\sigma}{2} (4m+h)$. 
The grand canonical potential per lattice site is computed at temperature $T$ as 
\be
\omega= - \frac{T}{N} \sum_{\vk \sigma} \log (1+ {\rm e}^{-\xi_{\vk \sigma} / T} ) + \frac{2m^2}{g} \,. 
\ee
By minimizing the potential with respect to $m$, we obtain the self-consistency equation
\be
m=\frac{g}{2N}\sum_{\vk \sigma} \sigma f(\xi_{\vk \sigma}) \,,
\label{self1} 
\ee
where $f(x)= ({\rm e}^{x/T} +1)^{-1}$ is the Fermi distribution function. 

In the presence of a magnetic field, the magnetization $m$ can  exhibit a jump, i.e., a metamagnetic phase transition. 
The end point of the metamagnetic phase transition is called a critical end point (CEP) and 
is given by the condition $\left. \frac{\partial h}{\partial m} \right|_{\mu, T} = 
\left. \frac{\partial^2 h}{\partial m^2} \right|_{\mu, T}=0$: 
\bea
&& \frac{1}{g} =- \frac{1}{N} \sum_{\vk \sigma} f'(\xi_{\vk \sigma}) \,, 
\label{self2}
\\
&&0=\frac{1}{N} \sum_{\vk \sigma} \sigma f''(\xi_{\vk \sigma}) \,.
\label{self3}
\eea
Here $f'$ and $f''$ mean the first and second derivative with respect to $\xi_{\vk \sigma}$. 
The CEP is determined by  solving the coupled equations Eqs.~(\ref{self1}), (\ref{self2}), and (\ref{self3}). 
In the limit of $m\rightarrow 0$ at $h=0$, \eq{self1} is reduced to \eq{self2}; the coupled equations 
Eqs.~(\ref{self2}) and (\ref{self3}) then determine a TCP at $h=0$.

We also study the entropy associated with a metamagnetic transition to get insight into a magnetocaloric effect. 
It is straightforward to compute the entropy from $S=-\frac{\partial \omega}{ \partial T} |_{\mu,h}$, 
\be
S=\frac{1}{N} \sum_{\vk \sigma} \log (1+ {\rm e}^{-\xi_{\vk \sigma} / T} ) + 
\frac{1}{TN} \sum_{\vk \sigma} \xi_{\vk \sigma} f(\xi_{\vk \sigma}) \,.
\label{S}
\ee

\section{Results}
In the present model [\eq{hamiltonian}] we have two parameters,  the next nearest-neighbor integral $t'$ and 
the interaction strength $g$, and three variables, temperature $T$, the chemical potential $\mu$, and a magnetic field $h$. 
The presence of $t'$ breaks a particle-hole symmetry, which generates rich phase diagrams as we shall show below.  
We choose $t'/t=0.35$ \cite{hase97,singh01}, which was employed to discuss  
a nearly ferromagnetic compound Sr$_3$Ru$_2$O$_7$ (Ref.~\cite{misc-mackenzie12}) in the context of 
the interplay of ferromagnetism and electronic nematic order \cite{yamase13}. 
If the nematic interaction is set to zero, the model in Ref.~\cite{yamase13} would be reduced to the present model. 
We shall also present results for $t'=0$ to highlight the importance of $t'$.  
As a value of $g$, we consider various choices. 
We scan all three variables $T$, $\mu$, and $h$, and establish the phase diagram of ferromagnetic and metamagnetic transitions. 
While $h$ is usually very small compared to the scale of $t$ in experiments, 
we wish to clarify the whole phase diagram by considering $h$ as a free variable, 
because  we are interested in getting insight into the qualitative feature of ferromagnetic and metamagnetic 
phase transitions by considering the simplicity of 
the present model [\eq{hamiltonian}]. 
Obtained phase diagrams are symmetric with respect to the axis of $h=0$ and thus we study only 
the region $h\geq 0$. We measure all quantities with the dimension of energy in units of $t$ and put $t=1$.

\subsection {Typical phase diagram}
\begin{figure}[ht]
\centering
\includegraphics[width=12cm]{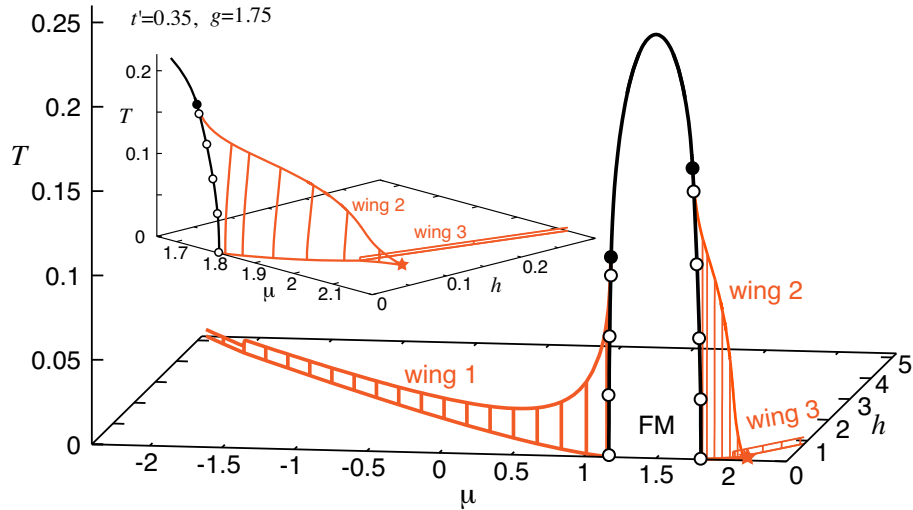}
\caption{Phase diagram in the $\mu$-$h$-$T$ space obtained for $t'=0.35$ and $g=1.75$. The ferromagnetic phase is realized in 
$1.17 \leq \mu \leq 1.81$ at $h=0$; the black line and the lines with open circles denote a second-order transition and first-order transitions, respectively, and the end points of the second-order transition line are TCPs (solid circles).  Two wings, wing~1 and wing~2, develop from the first-order transition lines upon applying $h$ and the wing~2 vanishes at a QCEP denoted by the star.  The wing~3 develops from the wing~2 at low $T$. The wings~2 and 3 are magnified in the inset. The upper edge of the wing corresponds to the CEL. The metamagnetic transition occurs by crossing the wings.   
}
\label{g7-wing}
\end{figure}

Figure~\ref{g7-wing} is the obtained phase diagram in the three-dimensional space spanned by $T$, $\mu$, and $h$. 
First we focus on the axis of $h=0$, where the phase diagram is shown in black. 
The ferromagnetic instability forms a dome-shaped phase transition line around van Hove filling $\mu_{\rm vH} = 4t' = 1.4$. 
The phase transition is of second order along the top of the dome and is of first order on the edges. 
The end points of the second-order transition line are TCPs.

When a field is applied, magnetization becomes finite and a ferromagnetic phase transition cannot be second order.  
Instead wing structures appear as shown in orange in \fig{g7-wing}. 
Two wings develop from the first-order transition lines at $h=0$ and describe first-order transition surfaces for a finite $h$. 
We refer to the wing on the side of the small (large) $\mu$ as the wing~1 (wing~2). 
In addition, another wing also develops from the wing~2 at low temperatures \cite{misc-wing3}, which we call the wing~3; 
the wings 2 and 3 are magnified in the inset of \fig{g7-wing}. 
These wings are located close to van Hove filling of either spin band in the presence of a field. 
Recalling the symmetry of $h \leftrightarrows -h$, six wings are realized in the $\mu$-$h$-$T$ space. 
The upper edge of the wing corresponds to a CEL, which is nothing less than a set of CEPs. 

\begin{figure}[ht]
\centering
\includegraphics[width=12cm]{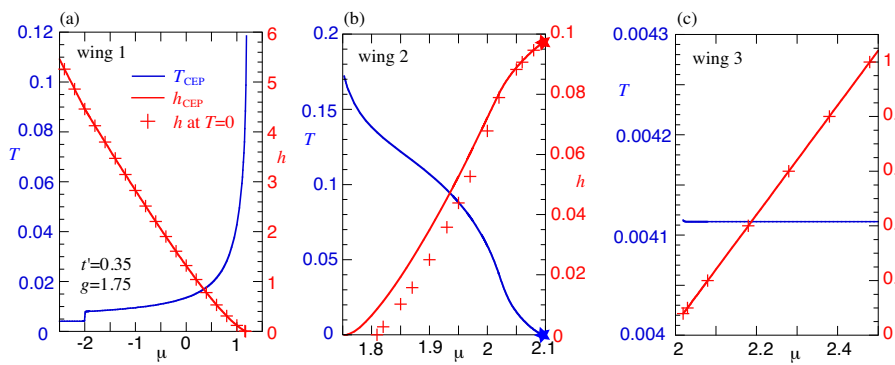}
\caption{Projection of the wing~1 (a), wing~2 (b) and wing~3 (c) in \fig{g7-wing} on the $\mu$-$T$ 
(left axis in blue) and $\mu$-$h$ (right axis in red) planes. $T_{\rm CEP}$ and 
$h_{\rm CEP}$ describe CELs and the symbol + denotes the position of the wing at $T=0$. 
The symbol star in (b) indicates a QCEP. 
$T_{\rm CEP} (=0.00411) $ in $\mu < -2$ in (a) is the same as that in $\mu \gtrsim 2$ in (c). 
}
\label{g7-phase}
\end{figure}

To see the wing structure more precisely, we project it on the $\mu$-$T$ plane and also 
$\mu$-$h$ plane in Figs.~\ref{g7-phase}(a)-(c) for the wing~1, 2, and 3, respectively. 
First we focus on \fig{g7-phase}(a). 
The projection on the $\mu$-$T$ plane indicates clearly that 
the critical end temperature is rapidly suppressed with decreasing $\mu$, 
exhibits a steep change at $\mu=-2$, and becomes constant. 
This anomaly at $\mu=-2$ indicates a Lifshitz point \cite{lifshitz60}, 
where the Fermi surface of the down-spin band disappears; see also \fig{FS}. 
It is interesting to see that the wing extends even beyond $\mu=-2$ without forming a QCEP. 
On the other hand, the projection on the $\mu$-$h$ plane shows that 
the critical end field ($h_{\rm CEP}$) is almost identical with a field of the metamagnetic transition at $T=0$, 
indicating that the wing stands almost vertically in the whole $\mu$-region. 
A close look at the curve of $h_{\rm CEP}$ reveals that the slope changes abruptly at $\mu=-2$ by crossing 
the Lifshitz point there; see also \fig{tp0-wing} where this change of the slope is more visible.

\begin{figure}[ht]
\centering
\includegraphics[width=7cm]{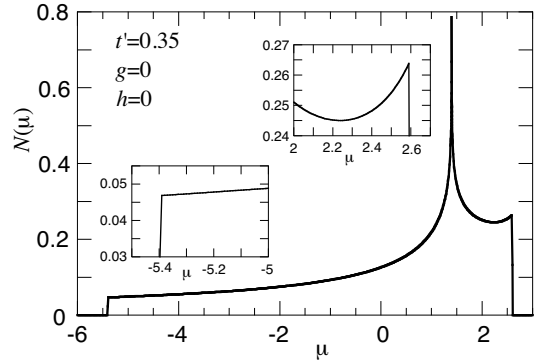}
\caption{Density of states for $t'=0.35$, $g=0$, and $h=0$. The insets magnify the regions close to the band edges.    
}
\label{DOS}
\end{figure}

These results are easily understood. 
The CEP is determined by Eqs.~(\ref{self1}), (\ref{self2}), and (\ref{self3}). 
Near van Hove filling, the DOS diverges logarithmically in the present model. 
Hence Eqs.~(\ref{self2}) and (\ref{self3}) 
can be fulfilled around van Hove filling. 
This means that the wing structure is realized close to van Hove filling in the $\mu$-$h$ plane, 
i.e., the $\mu$-$h$ curve in \fig{g7-phase}(a) traces van Hove filling of the up-spin band 
in the presence of the field. Let us define the {\it effective} chemical potential as 
\bea
\mu_{\uparrow} = \frac{4m+h}{2} + \mu \,,  
\label{muup} \\
\mu_{\downarrow} = -\frac{4m+h}{2} + \mu  
\label{mudown}\,.
\eea
When the up-spin density is at van Hove filling, namely $\mu_{\uparrow} = \mu_{\rm vH} = 4t'$, 
we obtain $\mu_{\downarrow} = 2 \mu - \mu_{\rm vH}$. On the other hand, the lower band edge is given by 
$\mu_{L} = -4 (t+t')$. Thus when $\mu_{\downarrow} < \mu_{L}$, the down-spin band becomes fully empty: 
see also the DOS in \fig{DOS}. 
This condition is given by $\mu < (\mu_{L} + \mu_{\rm vH})/2 = -2t = -2$, indicating that 
$\mu=-2$ corresponds to a Lifshitz point \cite{lifshitz60}. 
As long as the up-spin band is close to van Hove filling, i.e., $\mu_{\uparrow} = \mu_{\rm vH}$ even in $\mu < -2$, 
Eqs.~(\ref{self2}) and (\ref{self3}) always have a solution at $\mu_{\uparrow}=\mu_{\rm vH}$ at the CEP, 
namely at $h_{\rm CEP} = 2 (\mu_{\rm vH} -\mu_{\rm CEP})-4m_{\rm CEP}$, 
where $\mu_{\rm CEP}$ and $m_{\rm CEP}$ are the chemical potential and the magnetization at the CEP, respectively. 
Since only the up-spin band is active around van Hove filling, $m_{\rm CEP}$ becomes constant. 
Similarly the critical end temperature also becomes constant to be $T_{\rm CEP}=0.00411$. 
Hence the CEL is described by $h_{\rm CEP} = -2 \mu_{\rm CEP} + {\rm const.}$ in $\mu<-2$. 
This means that a QCEP cannot be present in a smaller $\mu$ region. 

Next we focus on \fig{g7-phase}(c). The projection on the $\mu$-$h$ plane shows that the CEL is located just above 
the line at $T=0$ with a good accuracy, that is, the wing~3 also stands vertically on the $\mu$-$h$ plane in \fig{g7-wing}. 
The wing~3 comes from the van Hove singularity  
of the down-spin band and the up-spin band is fully occupied. 
Hence the nature of a CEP is exactly the same as that in $\mu < -2$ in \fig{g7-phase}(a).  
In fact, $T_{\rm CEP}$ is the same value, there is no QCEP, and a CEP is described by 
$h_{\rm CEP} =  2 (\mu_{\rm CEP} - \mu_{\rm vH}) - 4 m_{\rm CEP}=  2 \mu_{\rm CEP}+ {\rm const.}$ 
The sign in front of $\mu_{\rm CEP}$ is opposite to the case in $\mu < -2$ because it is the down-spin band 
that fulfills van Hove filling for a large $\mu$. 
The wing~3 ends at $\mu=2.02$ by touching the wing~2 \cite{misc-wing3} as shown in Fig.~\ref{g7-wing}. 
The proximity to the wing~2 yields a tiny enhancement of $T_{\rm CEP}$ of the wing~3 there, which is barely visible 
in Fig.~\ref{g7-phase}(c). 

Details of the wing~2 are shown in \fig{g7-phase}(b). 
The curve projected on the $\mu$-$h$ plane indicates that in contrast to the case of the wing~1 and wing~3, 
the wing~2 is tilted to the large $h$ side, especially in a small $h$ region, although it may not be so  
clear in \fig{g7-wing}. The temperature dependence of the CEL is characterized by two different $\mu$ regions. 
For $1.75 \lesssim \mu \lesssim  2.0$, the effect of the van Hove singularity of the down-spin band 
is dominant, but the up-spin band is still involved in the DOS near the Fermi energy. 
This contribution is sizable because of the proximity to the band edge singularity. 
See the DOS in \fig{DOS} with a caveat that it is computed for $h=0$ and 
the effective chemical potential is given by Eqs.~(\ref{muup}) and (\ref{mudown}) in the presence of $h$. 
For $2.0 \lesssim \mu \lesssim 2.1$, the effect of the up-spin band becomes crucially important and we obtain two solutions 
satisfying Eqs.~(\ref{self1}), (\ref{self2}), and (\ref{self3}). 
One solution is obtained slightly below the up-spin band edge due to thermal broadening effect. 
With decreasing $T$, the solution is obtained closer to the up-spin band edge and exactly there at $T=0$, 
leading to a QCEP.  
Since the Fermi surface of the up-spin band disappears at the QCEP, it is a Lifshitz point \cite{lifshitz60}. 
The other solution is obtained when the up-spin band is fully occupied and the effective chemical potential is located at 
van Hove filling of the down-spin band, yielding the wing~3 in \fig{g7-wing} as we have already explained.  

\begin{figure}[ht]
\centering
\includegraphics[width=7cm]{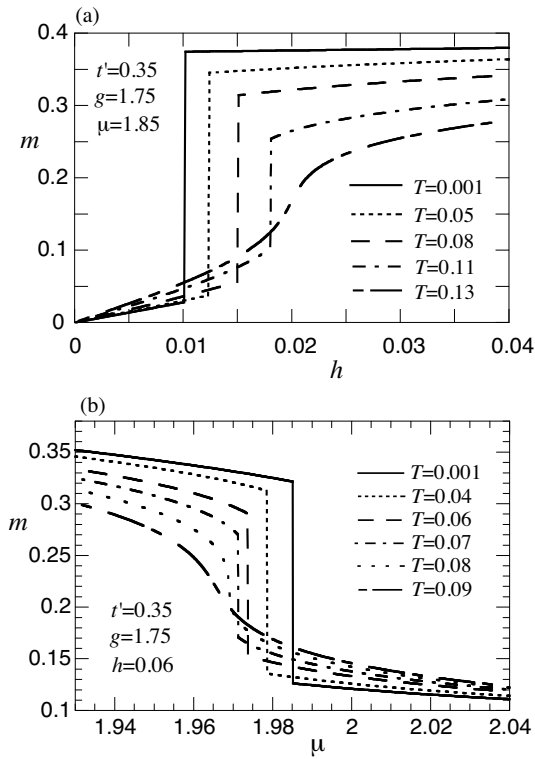}
\caption{(a) Magnetization $m$ as a function of a magnetic field $h$ (a) and the chemical potential $\mu$ (b) 
for several choices of temperatures. 
A metamagnetic transition occurs below $T=0.122$ (a) and $T=0.079$ (b) by crossing 
the wing. 
}
\label{meta}
\end{figure}

When crossing the wing, the magnetization exhibits a jump, namely a metamagnetic phase transition 
as shown in \fig{meta}(a). 
The jump of the magnetization is large at low temperature, becomes smaller with increasing $T$, 
and vanishes at the temperature of the CEP, which corresponds to the edge of the wing in \fig{g7-wing}. 
The magnetic field at which the jump occurs shifts to a high field with increasing temperature in \fig{meta}(a). 
This is because the wing~2 is tilted as shown in Fig.~\ref{g7-phase}(b). 
Although a metamagnetic transition is usually envisaged as a first-order transition as a function of a field, 
the jump of the magnetism also occurs as a function of the chemical potential (density) for a fixed field 
as shown in \fig{meta}(b). 
In this sense the wing structure in \fig{g7-wing} describes a {\it generalized} metamagnetic transition. 
The $\mu$ scan does not cross the wing~2 above $T=0.079$, where we have a smooth change of $m$, 
i.e., a crossover between the large and small magnetization.

\begin{figure}[ht]
\centering
\includegraphics[width=7cm]{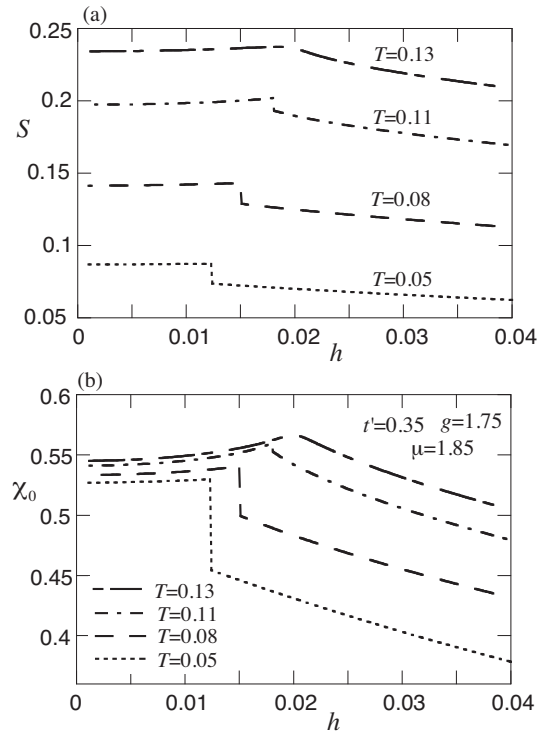}
\caption{(a) Entropy as a function of a field $h$ for several choices of temperatures  
for the same parameter set as those in \fig{meta}(a). (b) Similar plot for $\chi_{0}$, which is defined as 
the spin-summed temperature-averaged DOS. 
 }
\label{entropy}
\end{figure}

\begin{figure}[ht]
\centering
\includegraphics[width=7cm]{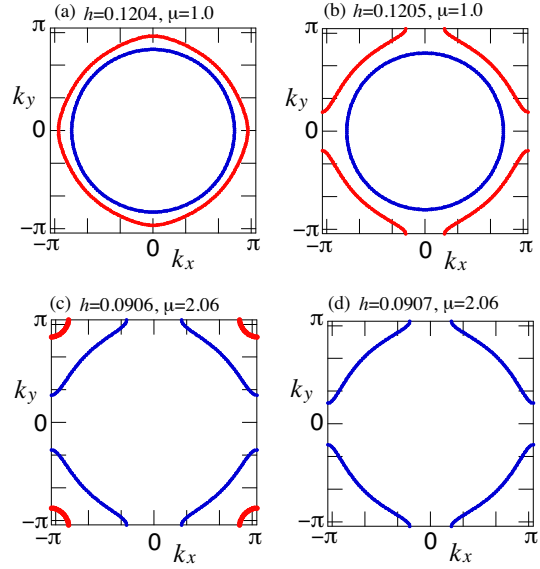}
\caption{Fermi surfaces of the up-spin (red) and down-spin (blue) band at $T=0.0002$ for $t'=0.35$ and $g=1.75$.  
(a) and (b) The topology of the up-spin Fermi surface changes by crossing the wing~1 in \fig{g7-wing}. 
(c) and (d) The up-spin Fermi surface disappears by crossing the wing~2 close to the QCEP in \fig{g7-wing}. 
 }
\label{FS}
\end{figure}

The entropy also exhibits a jump upon crossing the wing.
The evolution of the entropy as a function of $h$ is shown in \fig{entropy}(a) with the same 
parameters as those in \fig{meta}(a); results at $T=0.001$ is not shown since the entropy becomes 
too small to be seen in the scale of \fig{entropy}(a). 
Since the magnetization increases with increasing $h$ at each $T$ in \fig{meta}(a), 
one might assume that the entropy should decrease with increasing $h$. 
However, a close look at \fig{entropy}(a) reveals that the entropy slightly {\it increases} upon approaching the metamagnetic 
transition field from below. 
This is easily understood intuitively. 
We define $\chi_{0} = - \frac{1}{N} \sum_{\vk \sigma} f'(\xi_{\vk \sigma})$, which is 
the spin-summed DOS averaged over an energy interval of order of temperature. 
This quantity $\chi_{0}$ should continue to increase until the metamagnetic field because 
the large DOS is necessary to have a metamagnetic transition in the present model. 
This is indeed confirmed by the explicit calculations of $\chi_{0}$ as shown in \fig{entropy}(b). 
Recalling  Boltzmann's principle, we may associate the evolution of the entropy with that of the DOS as a function of $h$, 
which explains intuitively the reason why the entropy 
is enhanced in the vicinity of the metamagnetic transition with increasing $h$ from below in \fig{entropy}(a).

The relation between the entropy jump and the magnetization jump at the metamagnetic transition 
is described by the Clapeyron equation \cite{callen}: 
\be
\Delta S = - \Delta M \frac{d h} {d T} \,,
\label{clapeyron}
\ee
where $\Delta S$ and $\Delta M = \Delta m / g$ are jumps of the entropy and the magnetization, respectively. 
The key quantity here is $\frac{d h} {d T}$, which is a derivative along the coexistence curve, is zero when the wing stands vertically on the $\mu$-$h$ plane, and becomes finite when the wing is tilted. That is, it describes how much the wing is tilted with respect to the $\mu$-$h$ plane. 
Consequently, the entropy jump always becomes very small upon crossing a wing when it stands almost vertically.  
Hence if we compute the entropy jump across the wing~1 and the wing~3 in \fig{g7-wing}, it becomes 
very small even at a finite $T$ 
although the magnetization jump is sizable. 
This is the reason why we have chosen the wing~2 to describe the entropy jump in \fig{entropy}. 

At zero temperature, the entropy change becomes zero because of the third law of thermodynamics, 
yielding no jump of the entropy 
even crossing the wings. This implies $\frac{d h} {d T}=0$ at $T=0$, that is, 
the wing stands vertically on the $\mu$-$h$ plane in the limit of $T=0$ in \fig{g7-wing}.

In the present model, ferromagnetic and metamagnetic transitions are usually accompanied 
by a Lifshitz transition \cite{lifshitz60} at least at low temperatures. 
This is because they occur close to the van Hove singularity as well as 
the band edge via a first-order transition at low temperatures. 
There are two different types of Lifshitz transitions:  
i) the topology of the Fermi surface changes or 
ii) a Fermi surface of either up- or down-spin band disappears. 
The former type is observed when the metamagnetic transition occurs close to van Hove filling.  
Representative results are shown in Figs.~\ref{FS}(a) and (b). 
The latter type is expected when the metamagnetic transition occurs close to the band edge 
as shown in Figs.~\ref{FS}(c) and (d).

The obtained phase diagram shown in \fig{g7-wing} is regarded as a typical one of the itinerant ferromagnetic system 
in the presence of a field.  Generic features may be summarized as follows---quantitative features, of course, depend on details. 
The ferromagnetic instability occurs around van Hove filling at $h=0$ with a dome-shaped transition line. 
Applying the field $h$, the wing develops from the first-order transition line at $h=0$ 
as shown in \fig{g7-wing}. 
There are two important chemical potentials $\mu=-2$ and $2$, independent of $t'$ and $g$. 
The value of $\mu=2$ $(-2)$ corresponds to the case, in which the down-spin (up-spin) band is located at van Hove filling  
and the effective chemical potential of the up-spin (down-spin) [see Eqs.~(\ref{muup}) and (\ref{mudown})] 
sits at the upper (lower) edge of the band. 
Hence either spin band becomes full or empty above $|\mu| \gtrsim 2$ and only the other spin band becomes active there. 
As a result, the temperature and the magnetization at the CEP  
become constant along the CEL in $| \mu | \gtrsim 2$, which is described by 
$h_{\rm CEP} = 2 |\mu_{\rm CEP}|  + {\rm const.}$ 
In the present two-dimensional model, the DOS (see \fig{DOS}) 
exhibits a jump at the band edges and thus a temperature of a CEP 
shows a rapid change at $| \mu | = 2$. 
In particular, when the DOS is enhanced near the band edge as seen in the large $\mu$ side in \fig{DOS},  
two wings are realized around $\mu=2$. One dominant wing develops from 
the first-order transition line at $h=0$ and vanishes near $\mu=2$, leading to a QCEP, and 
the other wing forms a rectangular shape in a region of high $h$ as shown in \fig{g7-wing}.  
From a viewpoint of the topology of the Fermi surface, $\mu=-2$ and the QCEP correspond to Lifshitz points, 
where the Fermi surface of the down- and up-spin band vanishes, respectively.

\begin{figure}[bt]
\centering
\includegraphics[width=7cm]{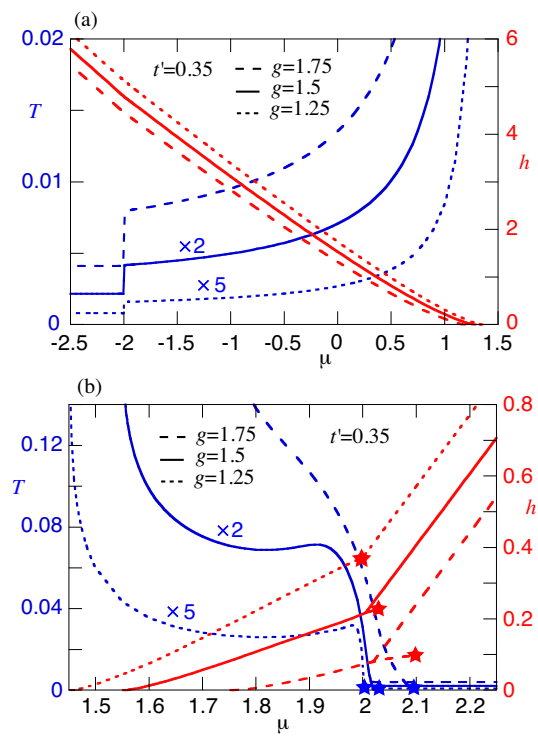}
\caption{Projection of the CELs on the $\mu$-$T$ (left axis in blue) and $\mu$-$h$ (right axis in red) planes 
for several choices of $g$ in the small (a) and large (b) $\mu$ region. $T_{\rm CEP}$ for $g=1.5$ and $1.25$ 
is multiplied by $2$ and $5$, respectively.  The CEL is not shown in the whole temperature region, but 
plotted only in a low-temperature region. 
The CEL in (a) corresponds to the wing~1. In (b) two CELs are drawn  for each $g$ and describe the wing~2 and 3, respectively. 
A QCEP (denoted by the star) 
associated with the wing~2 is realized at $\mu_{\rm QCEP}=2.095, 2.021$, and $2.003$ for $g=1.75, 1.5$, 
and $1.25$, respectively. 
$T_{\rm CEP}$ associated with the wing~3 is the same as that in $\mu < -2$; 
$T_{\rm CEP}=0.0041, 0.0011$ and $0.00016$ for $g=1.75, 1.5$, and $1.25$, respectively. 
}
\label{g-CEP}
\end{figure}

It might seem odd that the value of $|\mu |=2$ does not depend on $g$ and a QCEP always appears close to $\mu=2$ 
for the positive $t'$. 
This is, however, reasonable because the underlying mechanism lies in the proximity to the band edge and the 
enhancement of the DOS near $\mu \approx 2$ owing to the presence of $t'$ as we have explained. 
To emphasize this important insight, we show in \fig{g-CEP} how the CEL depends on $g$ by projecting it on 
the $\mu$-$T$ plane in blue. As expected, the temperature scale as well as the region of the ferromagnetic phase 
is shrunk for a smaller $g$ with keeping the qualitative features unchanged: 
the ferromagnetic phase is realized in $1.17 \leq \mu \leq 1.81$, $1.28 \leq \mu \leq 1.57$, 
and $1.35 \leq \mu \leq 1.46$ with the maximal $T_c$ around $0.25$, $0.12$, and $0.046$ for 
$g=1.75, 1.5$, and $1.25$, respectively. 
However, the value of $|\mu| = 2$ works as a {\it fixed} point because it describes the band edge when either spin band 
is located at van Hove filling in the presence of a field $h$. 
Hence a QCEP is always realized close to $\mu \approx 2$ and the constant temperature of the CEL appears 
in $|\mu|  \gtrsim 2$ as shown in Fig.~\ref{g-CEP}. It is interesting to note that $T_{\rm CEP}$ slightly 
increases close to $\mu=2$ in \fig{g-CEP}(b) before vanishing at a QCEP for $g=1.5$ and $1.25$. 
In \fig{g-CEP} we also present the projection of the CEL on the $\mu$-$h$ plane in red. 
As we have already explained, the CEL is described by 
$h_{\rm CEP} = - 2 \mu_{\rm CEP}+ {\rm const.}$ in $\mu < -2$ and 
$h_{\rm CEP} =  2 \mu_{\rm CEP}+ {\rm const.}$ in $\mu >2$, and thus 
the slope does not depend on $g$ there. Around $\mu=2$ two CELs touch to each other because 
the wing~3 develops from the wing~2 at low temperature as seen in \fig{g7-wing}. 
In particular, when $g$ becomes small, e.g., $g=1.25$ in \fig{g-CEP}(b), 
the wing~3 is realized very close to the QCEP associated with the wing~2. 
Consequently, the QCEP seems to extend as a {\it line} along the wing~3, but $T_{\rm CEP}$ is 
not zero along the wing~3. 

\subsection{Phase diagram hosting two ferromagnetic regions} 
We have studied generic features of the ferromagnetic and metamagnetic transitions in the itinerant electron system by taking a relatively small $g$. Some special, but interesting features are realized for a larger $g$, which we present here and also in the next subsection. 

\begin{figure}[ht]
\centering
\includegraphics[width=12.1cm]{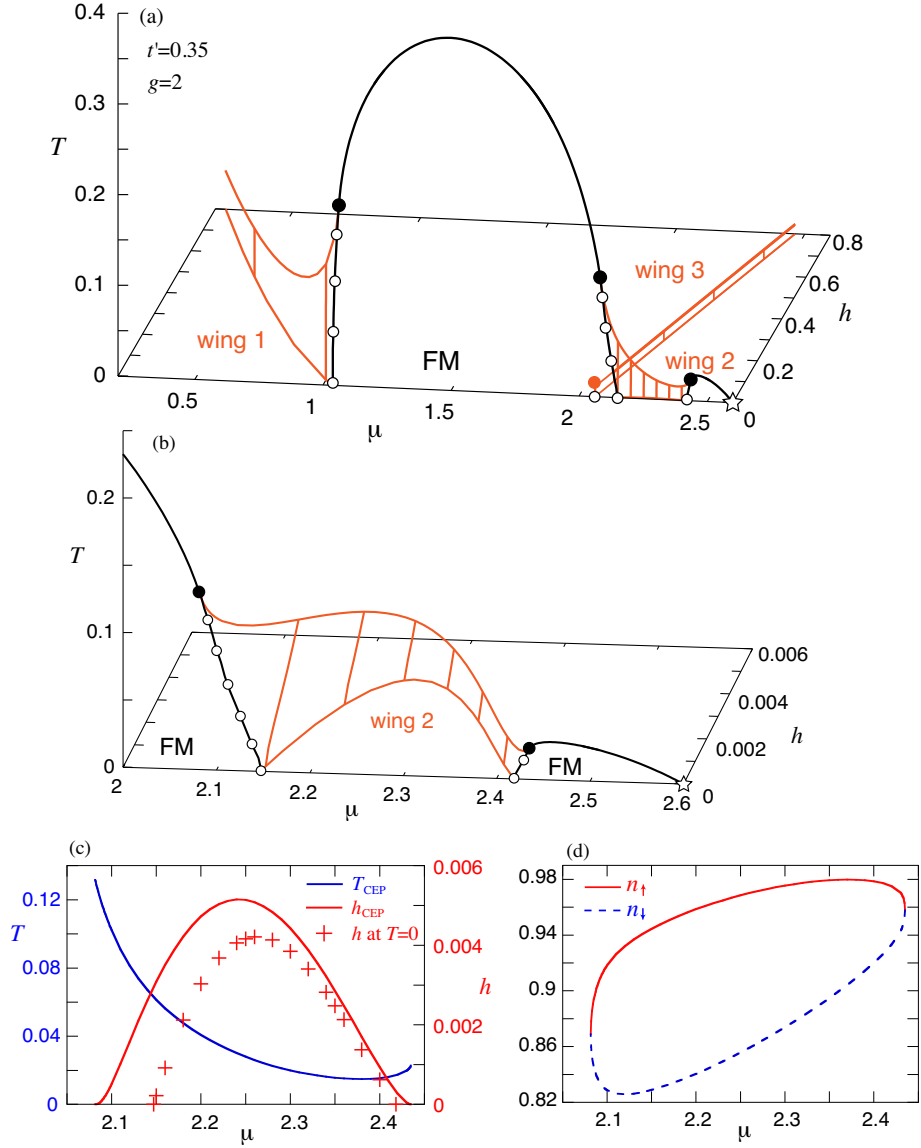}
\caption{\scriptsize{(a) Phase diagram in the $\mu$-$h$-$T$ space obtained for $t'=0.35$ and $g=2$. Two ferromagnetic regions are realized in 
$1.04 \leq \mu \leq 2.15$ and  $2.42 \leq \mu \leq 2.6$ at $h=0$; 
the black line and the lines with open circles denote a second-order transition and 
first-order transitions, respectively, and solid circles are TCPs and the star is a QCP. 
Inside the ferromagnetic phase, a metamagnetic transition occurs at $\mu=2.06$ and an orange circle denotes a CEP at $h=0$.  
The wing~1 develops from the first-order transition line upon applying $h$ on the small $\mu$ side and 
the wing~3 develops from the metamagnetic transition line at $h=0$. 
The first-order transitions of the two ferromagnetic regions are connected with the wing~2 through a finite $h$ region. 
The wing~2 is magnified in (b). 
(c) Projection of the wing~2 on the $\mu$-$T$ and $\mu$-$h$ planes. 
$T_{\rm CEP}$ and $h_{\rm CEP}$ describe the CEL and the symbol + denotes the position of the wing at $T=0$. 
(d) Density of the up- ($n_{\uparrow }$) and down-spin ($n_{\downarrow }$) along the CEL of the wing~2.}
}
\label{g8-wing}
\end{figure}

Typically one ferromagnetic phase is realized around van Hove filling at $h=0$ as we have already shown in \fig{g7-wing}. 
However, there is a parameter, which yields an additional ferromagnetic region at $h=0$ close to the band edge. 
As a representative result we choose $g=2$ and show the obtained phase diagram in \fig{g8-wing}(a); 
a region close to the band edge is magnified in \fig{g8-wing}(b). 
On top of a ferromagnetic region around van Hove filling ($\mu=1.4$), 
an additional ferromagnetic region emerges in $2.42 \leq \mu \leq 2.6$ at $h=0$, 
with a first-order transition near $\mu=2.4$ and a quantum critical point (QCP) at $\mu=2.6$.  
This phase is due to the band edge singularity of the DOS as shown in \fig{DOS}, where 
the DOS slightly increases toward the band edge. 

In this setup, a magnetic field yields a wing, referred to as the wing~2, which bridges 
two first-order transitions at $h=0$ through a finite $h$ region. 
The scale of the magnetic field is very small, indicating that the wing is realized very close to the $h=0$ line. 
In contrast to the case in \fig{g7-wing}, no QCEP is realized. 
To clarify this new wing structure, we project it on the $\mu$-$T$ planes as shown in Fig.~\ref{g8-wing}(c). 
Interestingly, $T_{\rm CEP}$ is not monotonic as a function of $\mu$. 
Figure~\ref{g8-wing}(c) also shows that the wing~2 is tilted substantially to the large $h$ side. 
In \fig{g8-wing}(d), we plot  the density of each spin along the CEL of the wing~2. 
The system is close to the band edge, and $n_{\uparrow}$ and $n_{\downarrow}$ have $2-13$\% 
and $4-17$\% holes, respectively. 

In \fig{g8-wing}(a) the metamagnetic transition line emerges at low temperatures 
inside the ferromagnetic phase at $\mu=2.06$ even at zero field. 
With applying a field, a wing structure develops with a rectangular shape. 
This wing originates from the van Hove singularity of the down-spin band and the underlying mechanism is 
the same as the wing~3 in \fig{g7-wing}. Therefore the CEL is described by 
$h_{\rm CEP} = 2\mu_{\rm CEP} + {\rm const.}$, and $T_{\rm CEP}= 0.0113$ independent of $h$. 

While the wing~1 is not shown in $\mu<0.4$ in \fig{g8-wing}(a), it extends to a smaller $\mu$ with lowering $T_{\rm CEP}$. 
Then $T_{\rm CEP}$ decreases rapidly around $\mu=-2$ and becomes  constant in $\mu<-2$, similar to 
the wing~1 in \fig{g7-wing}. 
$T_{\rm CEP}$ in $\mu \leq -2$ becomes 
the same temperature as $T_{\rm CEP}$ of the wing~3 in \fig{g8-wing}(a). 
This is because in both cases the CEP is determined by the van Hove singularity of either spin band and the other spin band 
is inactive. 

\subsection{Phase diagram for large interaction strength} 
\begin{figure}[ht]
\centering
\includegraphics[width=9cm]{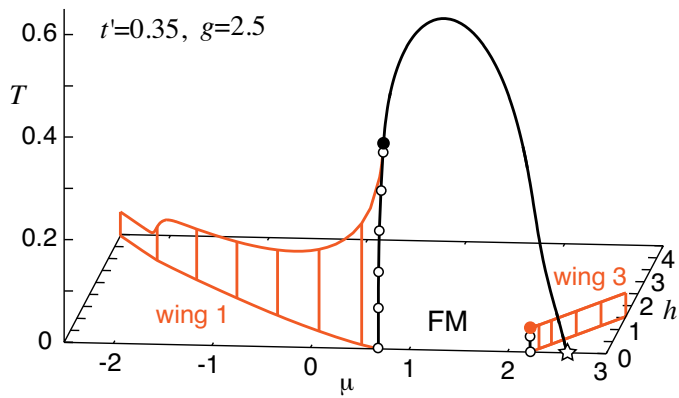}
\caption{Phase diagram in the $\mu$-$h$-$T$ space obtained for $t'=0.35$ and $g=2.5$. The ferromagnetic phase is realized in 
$0.69 \leq \mu \leq 2.6$ at $h=0$; the black line and the lines with open circles denote a second-order transition and 
first-order transitions, respectively; the solid circle is a TCP and the star a QCP. 
The metamagnetic transition occurs even inside the ferromagnetic phase in low $T$ at $\mu=2.23$  
and the orange circle denotes a CEP at $h=0$. 
Two wings develop from the first-order line and the metamagnetic transition line upon applying $h$, and are 
depicted by the wing~1 and wing~3, respectively. The wing~3 has  $T_{\rm CEP}=0.0465$, 
the same value as that of the wing~1 in $\mu \lesssim  -2$. 
}
\label{g10-wing}
\end{figure}

The ferromagnetic regions marge into the major phase for larger interaction strength. 
Consequently, as shown in \fig{g10-wing}, 
a single ferromagnetic phase is realized at $h=0$ 
with a second-order transition on the large $\mu$ side, yielding a ferromagnetic QCP 
at the band edge at $\mu=2.6$. Inside the ferromagnetic phase, the metamagnetic transition line is realized 
at $\mu = 2.23$ even without a field, similar to the case at $\mu=2.06$ in \fig{g8-wing}(a). 
Upon applying a field, a wing develops from the first-order transition line around $\mu \approx 0.7$ at $h=0$. 
This wing is qualitatively the same 
as the wing~1 in Figs.~\ref{g7-wing} and \ref{g8-wing}(a): $T_{\rm CEP}$ shows a rapid change around $\mu=-2$ 
and becomes constant to be $T_{\rm CEP}=0.0465$ in $\mu < -2$, where only the up-spin band is active. 
In contrast to Figs.~\ref{g7-wing} and \ref{g8-wing}, a wing corresponding to the wing~2 is not realized. 
The wing~3 is essentially the same as that in \fig{g8-wing}(a) and develops from the metamagnetic transition line at $h=0$, 
forming a rectangular shape, where only the down-spin band is active and thus $T_{\rm CEP}$ becomes equal to 
that of the wing~1 in $\mu< -2$. 
Since both wing~1 and wing~3 originate from the van Hove singularity of either spin band,  
no QCEP is realized even in a large $h$ region. 

\subsection{Particle-hole asymmetry} 
The presence of a finite $t'$ is responsible for the rich phase diagrams shown in Figs.~\ref{g7-wing}, \ref{g8-wing}, and \ref{g10-wing}. 
A large asymmetry with respect to van Hove filling 
and a metamagnetic transition even inside the ferromagnetic phase shown in Figs.~\ref{g8-wing} and \ref{g10-wing}  
are in fact owing to a finite $t'$. 
In particular, $t'$ yields a slight enhancement of the DOS 
near the band edge on the large $\mu$ side in \fig{DOS}. 
This seemingly subtle effect gives a large impact on the phase diagram of the ferromagnetic and metamagnetic transitions  
such as  a wing terminating with a QCEP for a finite field (\fig{g7-wing}), 
an additional ferromagnetic region close to the band edge (\fig{g8-wing}), 
and a wing connecting to two first-order transition lines (\fig{g8-wing}).

While we have presented results for $t'>0$, similar phase diagrams would be obtained for $t'<0$, where 
a wing structure on the large (small) $\mu$ side in Figs.~\ref{g7-wing}, 
\ref{g8-wing}, and \ref{g10-wing} are realized on the small (large) $\mu$ side. 
Hence a QCEP would be realized close to $\mu=-2$.

\subsection{Phase diagram for $t'=0$} 
\begin{figure}[tb]
\centering
\includegraphics[width=12cm]{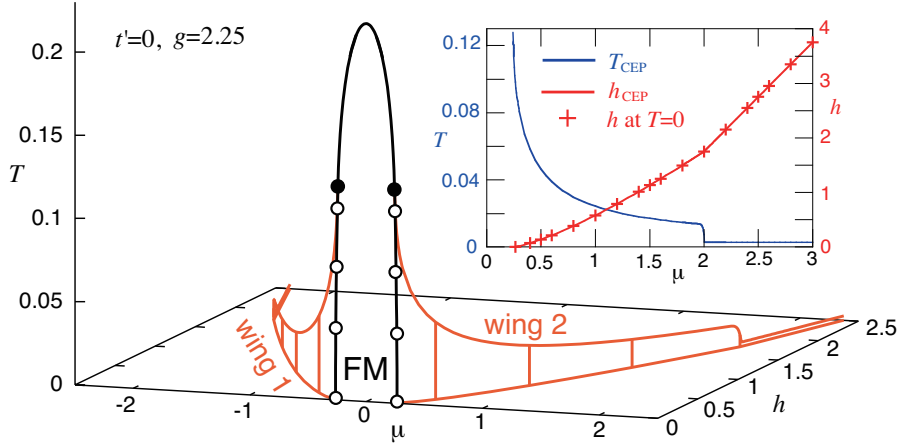}
\caption{Phase diagram in the $\mu$-$h$-$T$ space obtained for $t'=0$ and $g=2.25$. The ferromagnetic phase is realized in 
$-0.26 \leq \mu \leq 0.26$ at $h=0$; the black line and the lines with open circles denote a second-order transition and 
first-order transitions, respectively, and solid circles are TCPs. 
Two wings, wing~1 and wing~2, develop from the first-order transition lines upon applying $h$.  
The phase diagram is symmetric with respect to $\mu=0$.  
The inset shows a projection of the wing~2 on the $\mu$-$T$ (left axis in blue) and $\mu$-$h$ (right axis in red) planes. 
$T_{\rm CEP}$ and $h_{\rm CEP}$ describe the CEL and the symbol + denotes the position of the wing at $T=0$. 
The CEL becomes straight in $\mu > 2$, where $T_{\rm CEP}=0.0028$.   
}
\label{tp0-wing}
\end{figure}

The importance of $t'$ is highlighted by results for $t'=0$ shown in \fig{tp0-wing}. 
The phase diagram becomes symmetric with respect to $\mu=0$ owing to a particle-hole symmetry. 
A ferromagnetic phase is realized with a dome shape at $h=0$ and the transition is of first order 
at low temperatures. Upon applying a field, a wing structure develops along van Hove filling of either spin band. 
The temperature of a CEP drops at $| \mu | =2$ and 
concomitantly, $h_{\rm CEP}$ also changes the slope at $| \mu |=2$ as seen in the inset 
in \fig{tp0-wing}, where the wing is projected on the $\mu$-$T$ and $\mu$-$h$ planes. 
The CEP is realized even in $| h | >2$ with a constant $T_{\rm CEP} (=0.0028)$, where 
only the either spin is active and the Fermi surface of the other spin band vanishes.  
Since $h_{\rm CEP}$ is practically the same as a metamagnetic field at $T=0$ as seen in the inset, 
the wing in \fig{tp0-wing} stands vertically on the $\mu$-$h$ plane. 

\begin{figure}[ht]
\centering
\includegraphics[width=7cm]{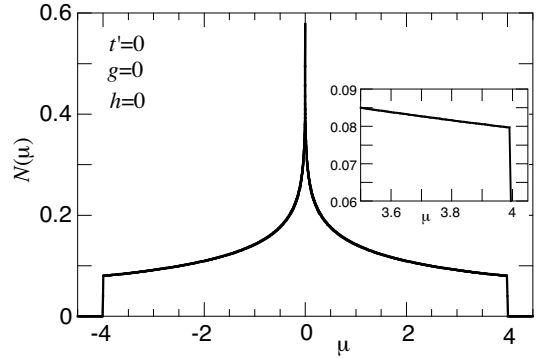}
\caption{Density of states for $t'=0$, $g=0$, and $h=0$. 
The inset magnifies a region close to the band edge. 
}
\label{tp0-DOS}
\end{figure}

In contrast to the cases in the presence of $t'$, 
a qualitatively different feature does not appear even for changing the interaction strength $g$ 
at least in $g \lesssim 5$. 
This is easily understood by noting a shape of the DOS. 
As shown in \fig{tp0-DOS}, the DOS exhibits the van Hove singularity at $\mu=0$ and 
decreases monotonically upon going away from van Hove filling. 
Even if the chemical potential approaches the band edge, 
no enhancement of the DOS occurs as shown in the inset. 
This feature is very similar to the one on the low $\mu$ side in the presence of $t'$ shown in \fig{DOS}. 
This explains the reason why the wing structures shown in \fig{tp0-wing} becomes similar to that on the low $\mu$ side 
shown in Figs.~\ref{g7-wing}, \ref{g8-wing}, and \ref{g10-wing}. 

\section{Conclusions and discussions} 
\subsection{Theoretical perspective}
The major point of the present work is to elucidate the actual phase diagrams 
of ferromagnetic and metamagnetic transitions in the three-dimensional space of 
the chemical potential $\mu$, a magnetic field $h$, and temperature $T$ by performing 
a microscopic study beyond phenomenological works so far \cite{levitin89,wohlfarth62,lidiard51,bean62,shimizu64,shimizu82,yamada93,goto01,yamada03,goto97,belitz05,yamada07,berridge10}. 
In particular, we have found that even a small enhancement of the DOS at the band edge leads to rich phase diagrams  
as shown in Figs.~\ref{g7-wing}, \ref{g8-wing}, and \ref{g10-wing} on the large $\mu$ side. 

While we have studied a square lattice system, our obtained results can form the foundation for itinerant 
ferromagnetic and metamagnetic transitions in other systems. 
As early theoretical studies \cite{levitin89,wohlfarth62,lidiard51,bean62,shimizu64, shimizu82} 
revealed, the large DOS with a positive curvature as a function of energy 
is crucial to ferromagnetic and metamagnetic transitions. 
In the square lattice system, such a condition is fulfilled around van Hove filling as well as the band edge 
in the presence of a finite $t'$. 
In other systems, the microscopic origin to fulfill that condition may become different. 
However, we expect that resulting phase diagrams including the wing structure may not 
differ essentially from the present results as long as the DOS near the Fermi energy plays a crucial role. 
In addition, while the physics we have described is based on the band structure, 
the band dispersion does not necessarily mean a bare one, but can be a renormalized one, for example, 
obtained after a mean-field-like approximation. 
One would consider multi-orbital degrees of freedom as well as spin-orbit coupling important in metamagnetic materials. 
Given that the present simple model already leads to the complex phase diagrams, a resulting phase diagram would become highly complicated especially when 
several bands cross the Fermi energy and yield a large DOS relevant to a metamagnetic transition. Even in this case, 
the present results would offer a solid basis to construct a complicated phase diagram. 
On the other hand, if practically only one band yielding a large DOS crosses the Fermi energy, 
a phase diagram similar to the present results would be 
obtained---the spin degrees of freedom in the present model may be then regarded as pseudospins  
describing the two Kramers degenerate states in zero magnetic field.

A physics different from ours was proposed. 
Reference~\cite{belitz99} showed that the low-energy particle-hole excitations, which are always present 
in a metallic system and independent of the band structure, generate a singular contribution to the free energy. 
As a result, they pointed out that a ferromagnetic order generally occurs via a first-order transition 
at low temperatures. 
Consequently, a wing structure also develops from the first-order transition line 
upon applying a field \cite{belitz05}.  
A resulting phase diagram becomes qualitatively similar to our results. 
There are, however, crucial differences, which may serve to identify the mechanism of 
ferromagnetic and metamagnetic transitions by considering the following: 
i) whether the magnetic transition is accompanied by a Lifshitz transition, 
ii) whether multiple wings are realized, and 
iii) whether a wing also develops even from the deeply inside of the ferromagnetic  phase. 
These possibilities may not occur in general in the scenario proposed in Ref.~\cite{belitz99} 
whereas they can occur in the band structure scenario. 
In the present model, 
the point~i) holds in general  at least at low temperatures, 
the point~ii) also holds close to the band edge with a finite $t'$, and 
the point~iii) is the case for a large interaction strength with $t'$.  
Those characteristic features in our model are shared to some extent with results 
obtained in a two-band Anderson lattice model \cite{wysokinski15}, 
where the DOS plays an important role of the metamagnetic transition, although 
the DOS contains a hybridization gap in contrast to our case (\fig{DOS}).

The wing structure developing from a first-order transition is not a special feature of the ferromagnetic system 
when applying a field. 
In the mixture of $^3$He-$^4$He, Griffiths pointed out the emergence of the wing structure from a first-order 
transition when applying a field conjugate to the superfluid density \cite{griffiths70}, although 
such a field cannot be controlled in experiments. 
In addition, similar wings were also reported 
by applying $xy$-anisotropy to a system where the electronic nematic instability 
occurs via a first-order transition in the tetragonal phase \cite{yamase15}.  
The $xy$-anisotropy can be controlled by uniaxial pressure and strain, and corresponds to a quantity 
conjugate to the nematic order parameter. 
The feature common to those three cases lies in that a TCP is present without a field and a wing is realized 
by applying a field conjugate to the order parameter. 
Referring to Griffiths's pioneering work about the wing structure, we may call it the Griffiths wing.

We remark on the magnetocaloric effect associated with a metamagnetic transition. 
Although a jump of magnetization occurs upon crossing the Griffiths wing in Figs.~\ref{g7-wing}, 
\ref{g8-wing}, \ref{g10-wing}, and \ref{tp0-wing}, this does not necessarily imply a sizable magnetocaloric effect. 
As seen in \eq{clapeyron}, a value of $\frac{d h}{d T}$ also does matter as was already pointed out in 
Refs.~\cite{yamada03} and \cite{fujita03}. 
It becomes nearly  zero as long as the Griffiths wing stands almost vertically 
on the $\mu$-$h$ plane and exactly zero at $T=0$. A sizable magnetocaloric effect is expected only around the wing~2 in the present minimal model, where the wing is tilted toward a higher field in Figs.~\ref{g7-wing} and \ref{g8-wing}. 
The present work, therefore, suggests that a value of $\frac{d h}{d T}$ along the coexistence line 
is more crucial than the jump of the magnetization associated with a metamagnetic transition 
to achieve a large magnetocaloric effect in metamagnetic materials.

How about the effect of ferromagnetic fluctuations on the phase diagram? 
Since we have employed a purely forward scattering model, fluctuations vanish in the thermodynamic limit.  
Hence we need to change the interaction term in \eq{hamiltonian} to, for example, one allowing a finite momentum transfer. 
An analysis of the Griffiths wing associated with the electronic nematic phase transition \cite{yamase15} 
showed that fluctuations beyond mean-field calculations suppress 
the tendency of a first-order transition and can wash out a part of the Griffiths wing at low temperatures. 
We therefore infer a similar case for the ferromagnetic and metamagnetic transition. 
The wing~2 in \fig{g7-wing} could be shrunk and the QCEP would be realized closer to the first-order ferromagnetic transition.  
This is an intriguing situation---strong fluctuations occur even near the first-order transition. 
Furthermore, the wing~1 could terminate with a new QCEP and the wing~3 could be fully washed out with reasonably 
strong fluctuations. 
These analyses including a resulting non-Fermi liquid state \cite{dzyaloshinskii96}, 
metamagnetic quantum criticality \cite{millis02}, and possible instabilities toward various phases such as 
triplet superconductivity \cite{katanin11}, spiral magnetism \cite{karahasanovic12}, 
a $d$-wave spin nematic order \cite{karahasanovic12}, and pair-density wave \cite{conduit13} 
are left to the future. 

\subsection{Experimental perspective} 
The effect beyond the present mean-field model is expected to be important to capture correctly the temperature dependence 
of the magnetic susceptibility, the specific heat, the nuclear spin relaxation rate, the resistivity, and 
others observed in experiments, as was extensively studied in the self-consistent renormalization theory \cite{moriya}. 
In addition,  given the minimal mean-field model [\eq{hamiltonian}], any quantitative comparison 
with experimental data is beyond the scope of the present work.
Therefore, we wish to discuss actual ferromagnetic systems focusing on 
a geometrical feature of the phase diagram. 
Moreover, although we have elucidated phase diagrams as a function of $\mu$, namely the electron density, 
our $\mu$ should not be viewed as a literal meaning when making a comparison with experiments. 
Rather it may be interpreted as a control parameter of the phase transition to tune the DOS at the Fermi energy. 
For example, if the pressure is a control parameter in experiments, 
we may assume that the pressure controls the evolution of the DOS at the Fermi energy.

UGe$_2$ is one of well studied ferromagnetic metals with strong spin anisotropy \cite{taufour10,kotegawa11,aoki12}. 
The transition is a first order at low temperatures, from which a wing develops. 
The wing was reported to disappear at a QCEP. This feature is similar to our wing~2 obtained in Fig.~\ref{g7-wing}. 
On top of that, a different wing also develops from a metamagnetic transition line inside the ferromagnetic phase 
(the boundary of FM1 and FM2 in Ref.~\cite{taufour10}), similar to our results 
around $\mu=2.06$ and $2.23$ in Figs.~\ref{g8-wing} and \ref{g10-wing}, respectively. 
While the present simple model cannot capture those two features 
for the same interaction strength, the present work suggests the importance of the enhancement 
of the DOS to understand the experimental data. 
The two wing structures in UGe$_2$ were also discussed in Ref.~\cite{wysokinski15} 
by invoking different variables in each case  in a two-band Anderson lattice model.

A wing structure was also reported in UCoAl \cite{aoki11}. This wing was extrapolated to terminate at a QCEP. 
Above a field of the QCEP, a metamagnetic crossover was reported. 
Such data may be interpreted in two different scenarios. 
One scenario is that the QCEP postulated in experiments is not actually present, but the temperature of a CEL 
simply becomes too low, not zero, to be detected. In the present model, temperature of a CEL 
drops abruptly when a band of either spin becomes empty or full; see the region 
around $\mu=-2$ in Fig.~\ref{g-CEP}(a), for instance. 
Another scenario is that the Griffiths wing indeed terminates at the postulated QCEP. The metamagnetic transition 
above the QCEP can be associated with another Griffiths wing developing from the vicinity 
of the QCEP as seen in \fig{g-CEP}(b). 
This temperature scale is, however, too low to be detected in experiments and thus 
the metamagnetism is observed as a crossover. 
In both scenarios, i) the postulated QCEP corresponds to a Lifshitz point \cite{lifshitz60} 
and ii) the metamagnetic crossover can be accompanied by a Lifshitz transition since the 
Griffiths wing is located close to van Hove filling, one of the typical features of a metamagnetic transition 
in terms of the large DOS.

ZrZn$_{2}$ \cite{kimura04,uhlarz04} is an itinerant ferromagnet and also exhibits a first-order transition at low temperatures. 
The above discussions about UGe$_{2}$ and UCoAl can also be applied to ZrZn$_{2}$ in the following points. 
First two metamagnetic transition lines associated with FM1 and FM2 are 
reported in Refs.~\cite{kimura04} and \cite{uhlarz04}, similar to UGe$_{2}$. 
In addition, the Griffiths wing was presumed to end at a QCEP, beyond which a metamagnetic crossover extends, similar to UCoAl. 

The metamagnetic transition inside the ferromagnetic phase observed in UGe$_{2}$ \cite{taufour10} 
and ZrZn$_{2}$ \cite{kimura04,uhlarz04} was discussed in terms of a double-peak structure of 
the electronic  DOS in Ref.~\cite{sandeman03}. In the present model, however, the metamagnetic transition inside the 
ferromagnetic phase around $\mu=2.06$ and $2.23$ in Figs.~\ref{g8-wing}(a) and \ref{g10-wing}, respectively, 
originates from a single-peak structure of the DOS, namely the van Hove singularity of the down-spin band inside 
the ferromagnetic phase. An insight similar to ours was also obtained in Ref.~\cite{wysokinski15}. 

Griffiths wings forming a double-wing structure were reported in LaCrGa$_{3}$ \cite{kaluarachchi17}. 
Although multiple Griffiths wings were also obtained in Figs.~\ref{g7-wing} and \ref{g8-wing} on the large $\mu$ side, 
the wing~2 and wing~3 are qualitatively different from the double-wing structure observed in LaCrGa$_{3}$. 
Considering that LaCrGa$_{3}$ also exhibits an antiferromagnetic phase next to the ferromagnetic phase, 
Ref.~\cite{belitz17} performed a phenomenological analysis by including both ferromagnetic and antiferromagnetic order 
parameters in the framework of the nonanalytic correction to the free energy from gapless particle-hole excitations. 
Obtained results successfully captured the overall phase diagram, but not the double-wing structure. 
It may be worthwhile to perform a microscopic analysis in a framework similar to the present work 
by including an antiferromagnetic interaction and 
see whether an important insight is obtained to understand the double Griffiths wings in  LaCrGa$_{3}$. 

UCoGe was reported to exhibit a first-order ferromagnetic transition at ambient pressure  \cite{ohta10}. 
If a TCP is hidden on the negative pressure side, we expect the emergence of a Griffiths wing 
upon applying a field, as suggested by  many theoretical studies \cite{griffiths70,lawrie84,belitz05,yamase15}. 
This possibility may be worth exploring further in experiments.  

In the experimental literature \cite{taufour10,kotegawa11,aoki11, kimura04,uhlarz04, kaluarachchi17}, 
the Griffiths wing is frequently assumed to end with a QCEP. 
This is supported theoretically by the mechanism of the nonanalytic correction from gapless particle-hole excitations 
around the Fermi surface \cite{belitz05} and also by a specific study of UGe$_{2}$ in a two-band Anderson lattice model \cite{wysokinski15}. On the other hand, from a view of the band-structure mechanism in the present simple model,  
the presence of a QCEP seems a delicate issue. While it is only the wing~2 in \fig{g7-wing} that 
terminates with a QCEP, the other wings extend up to a high field at very low temperatures. 
Moreover, as discussed in the last paragraph in Sec.~4.1, 
the wing~1 might also yield a QCEP by order-parameter fluctuations. 

Theoretical efforts \cite{konno91,evans92,ono98,meyer01,satoh01} were made to understand 
the metamagnetic transition observed in ${\rm CeRu_2Si_2}$ in terms of the periodic Anderson model, where 
conduction electrons hybridize with the almost localized $f$ electrons. 
We, however, point out that ${\rm CeRu_{2}Si_{2}}$ provides a setup very different from the present model 
in that it is a paramagnet characterized by antiferromagnetic correlations via the RKKY interaction \cite{kadowaki04,knafo09}, 
not ferromagnetic fluctuations in the vicinity of a ferromagnetic phase.  

We also provide a few remarks on ${\rm Sr_{3}Ru_{2}O_{7}}$, 
because it exhibits a wing structure in the space of a magnetic field, tilting angle of the field,  
and temperature \cite{misc-mackenzie12}. 
In the early days, ${\rm Sr_{3}Ru_{2}O_{7}}$ was envisaged as a system of a metamagnetic transition. However, later the dome-like phase was discovered as a function of a field, with a first-order transition at the edges of the dome. This cannot be interpreted as a usual metamagnetic phenomenon. Rather a $d$-wave Pomeranchuk instability \cite{yamase00,metzner00}, i.e., the electronic nematic instability was proposed. Since the nematic transition usually occurs via a first-order transition at low temperatures  \cite{khavkine04,yamase05}, the nematic transition as a function of a field is necessarily accompanied by a jump of the magnetization. Furthermore, the recent neutron scattering experiment \cite{lester15} revealed that the nematic dome is actually the dome of an incommensurate spin-density-wave (SDW) phase with a wavevector $\vq=(q_{x}, 0, 0)$, which is in line with a Ginzburg-Landau theory \cite{berridge10}. It is not yet settled which is the driving force to yield the dome-like phase, the nematic instability or the SDW instability or others. More works are necessary to provide additional insight into the wing structure observed in ${\rm Sr_{3}Ru_{2}O_{7}}$.

We conclude the present work with a future perspective to understand a metamagnetic transition in metals. 
It is the large DOS which drives a metamagnetic transition in the present model. 
It is, however, likely true that effects other than the large DOS may also play a certain role 
in actual materials, for example, nonanalytic corrections from gapless particle-hole excitations around 
the Fermi energy  \cite{belitz99} and magnetoelastic coupling \cite{oliveira05}.  
In particular, the former scenario was discussed for various itinerant ferromagnets \cite{karahasanovic12,conduit13,belitz05}. 
The latter effect was invoked to understand the large magnetocaloric effect in ${\rm MnFeP_{0.45}As_{0.55}}$ \cite{oliveira05}. 
In reality, these effects as well as the DOS are expected to work constructively to yield a metamagnetic transition. 
Hence the crucial issue is to identify which effect is actually dominant. 
The present work has clarified details of itinerant metamagnetism microscopically from a viewpoint of the large DOS and will 
serve as a solid basis toward more complete understanding of a metamagnetic transition in metallic compounds. 

\section*{Acknowledgments}
The author thanks K. Kuboki for a critical reading of the manuscript and thoughtful comments. 
The author is also indebted to W. Metzner for stimulating discussions at the initial stage of the present work and to 
the warm hospitality of Max-Planck-Institute for Solid State Research. 
This work was supported by JST-Mirai Program Grant Number JPMJMI18A3, Japan and JSPS KAKENHI Grants No.~JP20H01856.

\section*{References}
\bibliography{main} 

\begin{thebibliography}{10}
\bibitem{misc-wing3}
The touch of the wing~3 with the wing~2 at $\mu=2.02$ shares the same feature
  with that of the wing~3 with the ferromagnetic phase in Fig.~8(a) and 9, for
  the ferromagnetic phase is regarded as a first-order transition surface
  separating the state with a positive $m$ in $h>0$ and a negative $m$ in
  $h<0$.
\end{thebibliography}
\end{document}